\documentclass[twocolumn,floatfix,prb]{revtex4-1}
\usepackage{graphicx}
\usepackage{bm}
\usepackage{color}
\usepackage{mathtools}
\usepackage{amsfonts}
\newcommand{\bk}{{\bf k}}
\newcommand{\bq}{{\bf q}}

\newcommand{\bz}{\mathfrak{z}}

\begin{document}
\title{Influence of the ferroelectric quantum critical point on SrTiO$_3$ interfaces}
\author{W. A. Atkinson} \email{billatkinson@trentu.ca}
\author{P. Lafleur}
\author{A. Raslan}
\affiliation{Department of Physics and Astronomy, Trent University,
  Peterborough Ontario, Canada, K9J7B8} 
\date{\today}
\begin{abstract}
We study a model SrTiO$_3$ interface in which conduction $t_{2g}$ electrons couple to the ferroelectric (FE) phonon mode.  We treat the FE mode within a self-consistent phonon theory that captures its quantum critical behavior, and show that proximity to the quantum critical point leads to universal tails in the electron density of the form $n(z) \sim (\lambda+z)^{-2}$, where $\lambda \sim T^{2-d/\bz}$, with $d=3$ the dimensionality and $\bz=1$ the dynamical critical exponent.  Implications for the metal-insulator transition at low electron density are discussed.
\end{abstract}
\maketitle

\section{Introduction}

SrTiO$_3$ (STO) is  remarkable for being a quantum paraelectric.\cite{Barrett:1952ww,Muller:1979wa}  Energetically, the lattice favors a ferroelectric (FE) distortion; however, this distortion is suppressed by quantum fluctuations.  The incipient FE transition is associated with a transverse optical phonon in which the O$^{2-}$ anions move oppositely to the Sr$^{2+}$ and Ti$^{4+}$ cations.\cite{Cowley:1962wt,Cowley:tr}  The phonon frequency $\omega_\bq$ softens dramatically at $\bq=0$ as the temperature $T$ is lowered, but saturates below $T \sim 15$ K due to quantum effects.  A finite-temperature FE transition can be obtained by substitution of $^{18}$O for $^{16}$O,\cite{Itoh:1999eq} and it is possible to reach the quantum critical point (QCP) at which the FE transition temperature is 0~K by tuning the $^{18}$O fraction.  Because of this, bulk STO has been studied as a model system for quantum critical phenomena.\cite{Dec:1999wh,Dec:2005cr,Palova:2009js,Rowley:2014bda}  

\begin{figure}[tb]
\includegraphics[width=\columnwidth]{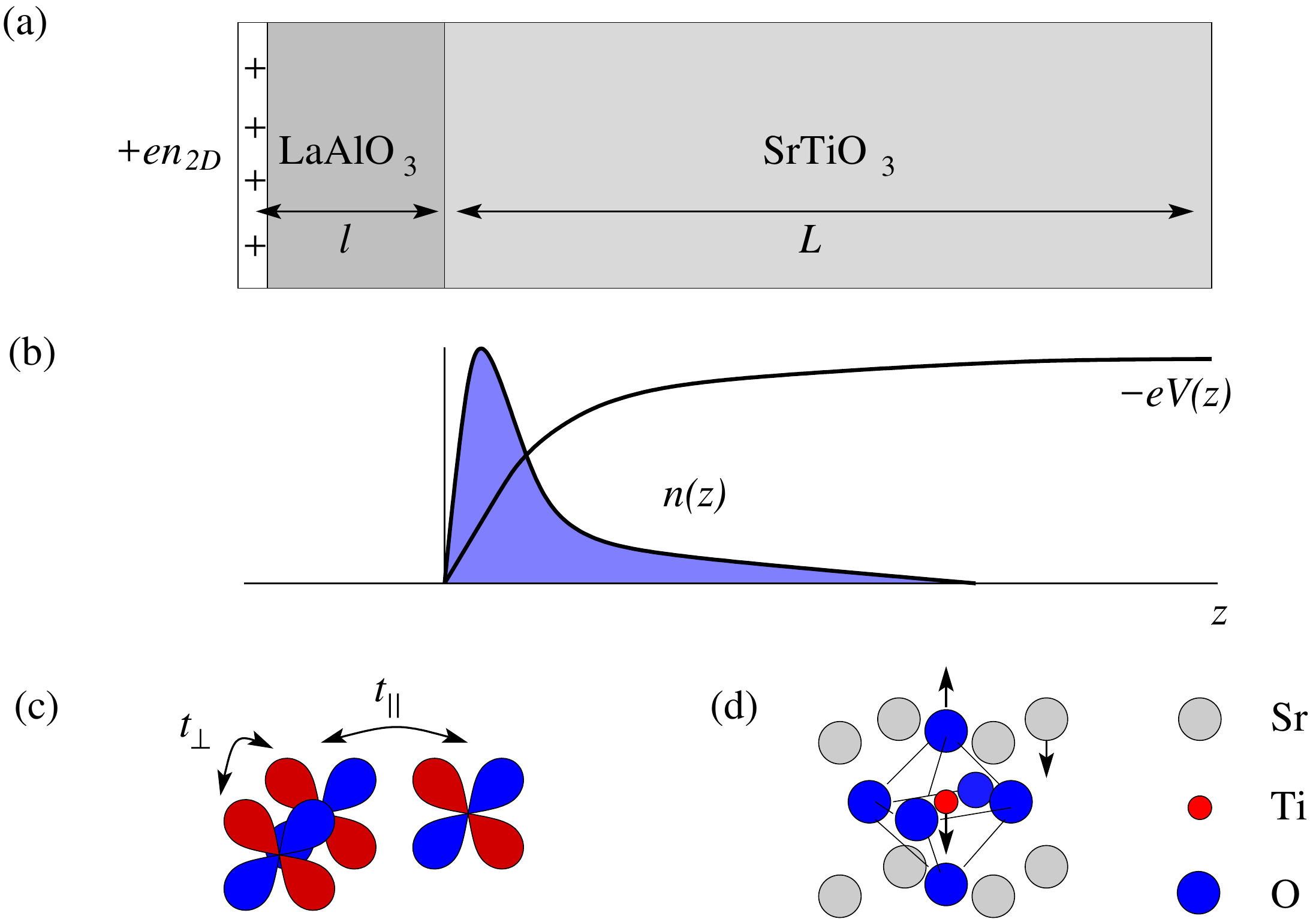}
\caption{(Color online)  (a) Structure of an n-type  LaAlO$_3$/SrTiO$_3$ interface.  To avoid a polar catastrophe, a two dimensional charge density $-en_\mathrm{2D}$ is transferred from the LaAlO$_3$ surface to the interface, leaving behind a residual positive surface charge $+en_\mathrm{2D}$.  (b) The residual LaAlO$_3$ surface charge creates a potential well $-eV(z)$ that confines the 2DEG to the STO side of the interface.  The electron density $n(z)$ in the SrTiO$_3$ has a strongly 2D component within $\sim 4$ nm of the interface and a tail that extends significantly farther.  (c) Conduction bands are formed from Ti $t_{2g}$ orbitals ($d_{xy}$, $d_{xz}$, and $d_{yz}$).  Electron hopping amplitudes between neighboring $t_{2g}$ orbitals of the same symmetry are $t_{\|}$ or $t_\perp$ depending on the orientation of the orbitals.   (d) Dielectric screening of electric fields occurs primarily through a soft phonon mode associated with incipient ferroelectric order.  In this mode, the Ti$^{4+}$ and Sr$^{2+}$ ions move oppositely to the octahedral oxygen cage, as indicated by the arrows.}
\label{fig:structure}
\end{figure}

Strontium titanate is also a key component of many oxide interfaces, and rose to prominence in this context following the seminal discovery by Ohtomo and Hwang that a two-dimensional electron gas (2DEG) forms at LaTiO$_3$/STO interfaces.\cite{Ohtomo:2004hma}  This was the first member of a growing  family of interfaces in which nonpolar STO is mated to one of several polar perovskites, most notably LaTiO$_3$, LaAlO$_3$, and GdTiO$_3$.  In this family, the bulk materials are insulating and the 2DEG forms on the STO side of the interface, as illustrated in Fig.~\ref{fig:structure}.  Ongoing interest in these interfaces is sustained by observations of  coexisting ferromagnetism and superconductivity,\cite{Brinkman:2007fk,Reyren:2007gv,Dikin:2011gl} nontrivial spin-orbit effects,\cite{BenShalom:2010kv,Caviglia:2010jv}  a metal-insulator transition,\cite{Liao:2011bk} and gate-controlled superconductivity.\cite{Caviglia:2008uh}  

Given the importance of quantum criticality in bulk STO, it is natural to ask how the 2DEGs at STO interfaces are affected by the FE QCP.  In many systems, inelastic scattering by critical fluctuations of the incipient order generates a power-law $T$-dependence in the quasiparticle lifetime which may be directly observed in transport experiments.\cite{Lohneysen:2007gm}  However, in STO the soft phonon mode associated with the FE transition has a transverse polarization and therefore couples only weakly to the conduction electrons;\cite{Ziman:1962} quantum criticality is therefore not easily observed in the transport properties.

Here, we show that quantum criticality has a profound effect on the 2DEG charge distribution at the interface.  In particular, the linear dielectric susceptibility $\chi_{\bq=0}$ is related to the $\bq=0$ phonon frequency $\omega_{\bq=0}$ by 
\begin{equation}
\chi_{\bq=0} = \frac{Q^2}{\epsilon_0 a_0^3 M\omega_{\bq=0}^2},
\end{equation}
where $Q$ and $M$ are the effective charge and mass for the FE phonon mode,  $a_0$ is the lattice constant, and $\epsilon_0$ is the permittivity of free space.  Near the QCP, $\omega_{\bq=0}$ decreases with decreasing temperature, leading to $\chi_{\bq=0}\sim 10^4$  at low $T$.  While it is understood that a large $\chi_\bq$ screens  interfacial electric fields and allows the 2DEG to spread into the STO at low temperatures, the connection to FE quantum criticality has not been explored.

The cartoon in Fig.~\ref{fig:structure} shows the case of a thin film of LaAlO$_3$ deposited on a slab of STO.  Because the LaAlO$_3$ is polar, a voltage difference proportional to its thickness builds up across the LaAlO$_3$ film.  To avoid large electrostatic energies (known as a ``polar catastrophe''), it is energetically favorable to transfer charge from the surface of the LaAlO$_3$ to the interface.\cite{Nakagawa:2006gt}  The amount  of transferred charge  can be as large as  $n_\mathrm{2D} = 0.5$ electrons per two-dimensional (2D) unit cell for the pure polar catastrophe case (as in GdTiO$_3$/STO interfaces), but is $\sim 10\%$ of this in most interfaces, likely because of oxygen vacancy formation at the LaAlO$_3$ surface during growth.\cite{Bristowe:2014fc,Yu:2014hx,Gariglio:2015jx}   $n_\mathrm{2D}$ can also be continuously adjusted by gating, down to a metal-insulator transition at $n_\mathrm{2D}\sim 0.01$ electrons per 2D unit cell.\cite{Liao:2011bk}  All three cases (polar catastrophe, surface vacancy mediated doping, and gating) can be modeled by a positive surface charge that confines a 2D charge density $-en_\mathrm{2D}$ on the STO side of the interface.

Guided by {\em ab initio} calculations,  simplified models of oxide interfaces have proven invaluable for understanding temperature effects and for studying large system sizes.\cite{Stengel:2011hy,Khalsa:2012fu,Zhong:2013cr,Reich:2015ut,Gariglio:2015jx,Raslan:2016} In Sec.~\ref{sec:calculations}, we describe our approach, which involves solving a set of coupled equations for the electronic density and the lattice polarization for the slab geometry shown in Fig.~\ref{fig:structure}.  The electronic calculations include the three relevant $t_{2g}$ Ti orbitals ($d_{xy}$, $d_{yz}$, and $d_{xz}$) that form the conduction bands, while the polarization calculations involve a self-consistent phonon calculation for the FE phonon mode.  This approximation includes both quantum and thermal fluctuations, and allows us to explore FE quantum criticality.  While conceptually straightforward, these calculations are complicated, and in Sec.~\ref{sec:results} we  use a simplified one-band model to interpret the results of our numerical calculations.  This simplified model admits analytic solutions, and allows us to demonstrate explicitly the impact of critical fluctuations on the interfacial charge distribution.

%Importance of lattice polarizability near interface.\cite{Hamann:2006hz}
%Core-shell models.\cite{BussmannHolder:2012df}
%
%Similar calculations were made previously,\cite{Prosandeev:1999tx,Salje:1991wu} however
%these calculations treated the mode as purely local, namely as an Einstein mode, so that there
%is no $\bq$-dependence.  This $\bq$-dependence is crucial for understanding interfaces.

\section{Calculations}
\label{sec:calculations}

We use a simple model first proposed by Schneider, Beck and Stoll\cite{Schneider:1976vn} for the ferroelectric phonon mode pictured in Fig.~\ref{fig:structure}(d).   In this model,  $\hat X_{j\alpha}$ denotes the operator form of the normal coordinate for the optical mode in unit cell $j$, with $\alpha = x,y,z$ the axis along which the unit cell is polarized.  The polarization of unit cell $j$ is then
\begin{equation}
P_{j\alpha} = \frac{Q\langle \hat X_{j\alpha} \rangle}{a_0^3},
\label{eq:P}
\end{equation}
where $Q$ is an effective charge for the optical mode and $a_0^3$ is the unit cell volume.  

The quantum Hamiltonian for the optical mode is taken to be
\begin{eqnarray}
\hat H_\mathrm{ph} &=& \sum_{i\alpha} \frac{\hat \Pi_{i\alpha}^2}{2M} 
+ \frac 12 \sum_{ij\alpha} \hat X_{i\alpha} D_{ij}^{0\alpha} \hat X_{j\alpha}  \nonumber \\
&& +\frac{B}{4\eta} \sum_i \left ( \sum_{\alpha} \hat X_{i\alpha}^2 \right )^2 - Q\sum_{i\alpha} E_{i\alpha} \hat X_{i\alpha} 
\label{eq:H}
\end{eqnarray}
where $\hat \Pi_{i\alpha}$ is the momentum operator conjugate to $\hat X_{i\alpha}$, satisfying
\begin{equation}
[\hat X_{i\alpha},\hat \Pi_{j\beta}] = i\hbar \delta_{i,j} \delta_{\alpha,\beta},
\end{equation}
$M$ is the optical mode effective mass, and $D^{0\alpha}_{ij}$ is the linear elastic constant between unit cells $i$ and $j$ for polarization in direction $\alpha$.  The parameter $\eta$ is the number of  polarization components (ie.\ $\alpha \in [1,\eta]$), and should be 3 in three dimensions. However, by treating $\eta$ as a fitting parameter, we obtain significantly better quantitative fits to both the field- and temperature-dependence of the measured susceptibility.\cite{Dec:1999wh}  The third term on the right hand side of Eq.~(\ref{eq:H}) describes nonlinear effects, while the final term couples the polarization to an electric field.  The parameters $Q$, $M$, and $B$, as well as the elastic constants $D^{0\alpha}_{ij}$ are determined by fitting to experimental measurements of the dielectric susceptibility and  phonon dispersion.  Model parameters are given in Table~\ref{table:1}.

\begin{table}
\begin{tabular}{c|c}
Parameter & Value \\
\hline\hline
$M$ & $4\times 10^{-26}$ kg \\
$Q$ & $5.4 e$ \\
$A$ & 0.004 eV \AA$^{-2}$  \\
$B$ & 160 eV \AA$^{-4}$ \\
$C$ & 0.375 eV \AA$^{-2}$ \\
$\eta$ & 12\\
\hline
\end{tabular}
\caption{Parameters for quantum phonon model.  See Appendix~\ref{app:A} for model details.}
\label{table:1}
\end{table}

Self-consistent phonon theory involves making a mean-field approximation for the nonlinear term in Eq.~(\ref{eq:H}) (Appendix~\ref{app:A}).  This approximation leads to a pair of self-consistent equations for the mean displacement $\langle \hat X_{i\alpha} \rangle$ [Eq.~(\ref{eq:X})] and the fluctuations $\langle \hat X^2_{i\alpha} \rangle  - \langle \hat X_{i\alpha} \rangle^2$ [Eq.~(\ref{eq:X2})].  These equations have been thoroughly discussed in the context of insulating SrTiO$_3$, and have been shown to generate critical behavior that is largely consistent with experiments.\cite{Prosandeev:1999tx,Palova:2009js,Rowley:2014bda}

The numerical calculations discussed in this section are for the slab geometry pictured in Fig.~\ref{fig:structure}.  We assume translational symmetry along the $x$ and $y$ directions (parallel to the interface) such  that the normal coordinate, polarization, and electric field are functions only of $z$ and are parallel to the $z$ axis, namely $\langle \hat X_{j\alpha} \rangle \rightarrow \langle \hat X \rangle_{j_z} \delta_{\alpha,z}$, etc.  For an STO slab of thickness $L$ unit cells, we adopt the boundary conditions that the polarization vanishes at the interface ($j_z=0$) and at the back of the slab ($j_z = L$).  The latter condition is motivated by the fact that the polarization vanishes in the bulk due to screening by the 2DEG; the boundary condition at $j_z=0$ assumes that the insulating side of the interface (eg.\ the LaAlO$_3$ film) is rigid and suppresses polarization of the top SrTiO$_3$ layer.  This assumption is not critical since, as we show below, the SrTiO$_3$ lattice relaxes within a few unit cells of the interface.

The electric field in Eq.~(\ref{eq:H}) depends on the polarization, the free electron density, and the charge density at the LaAlO$_3$ surface. It is obtained by solving a discrete version of Gauss' law
%\begin{equation}
%\epsilon_\infty \nabla\cdot {\bf E} = \rho - \nabla\cdot {\bf P},
%\end{equation}
to obtain the electrostatic potential\cite{Raslan:2016} 
\begin{equation}
V_{j_z} = -\frac{e}{2\epsilon_\infty a_0} \sum_{i_z} (|i_z-j_z|-i_z) (n^b_{i_z} - n_{i_z} + n_\mathrm{2D}\delta_{i_z,0}),
\label{eq:V}
\end{equation}
where $en^b_{i_z} = -\partial P /\partial z$ is the bound charge density, $\epsilon_\infty=5\epsilon_0$ is the optical dielectric constant, and $n_{i_z}$ is the free electron density in layer $i_z$ of the STO.

To obtain $n_{j_z}$, we solve the Schr\"odinger equation for the electronic wavefunctions for each of the $t_{2g}$ orbital symmetries, subject to the boundary condition that the wavefunctions vanish at $j_z=0$ and $j_z = L$ (Appendix~\ref{app:C}).  Electron-electron interactions are treated within a self-consistent Hartree approximation.  Because the $t_{2g}$ Wannier orbitals are localized about individual Ti atoms, a tight binding description of the electronic bands is appropriate.\cite{Stengel:2011hy,Zhong:2013cr}  In this description, there are two significant hopping matrix elements between adjacent orbitals of type $\alpha$ [Fig.~\ref{fig:structure}(c)]:  $t_\| \approx 235$ meV between Ti atoms in the plane of the orbital $\alpha$; and $t_\perp \approx 35$ meV between Ti atoms perpendicular to orbital $\alpha$.
This large anisotropy leads to an important distinction between bands with $xy$ symmetry and those with $xz$ and $yz$ symmetry: the effective mass for $xy$ bands is light in the $x$-$y$ plane and heavy along the $z$ direction, while both the $yz$ and $xz$ bands are light along the $z$ direction.  For typical electron densities, the lowest-energy $xy$ bands are therefore confined to within a few unit cells of the interface while the lowest-energy $yz$ and $xz$ bands extend several nm away from the interface.\cite{Delugas:2011ih}
    
We remark that we have neglected spin-orbit coupling in our calculations.\cite{Zhong:2013fv,Khalsa:2013hk}  As the region of interest for quantum criticality lies away from the interface, Rashba spin-orbit coupling is likely to be irrelevant to our discussion.  On the other hand, there is an atomic spin-orbit coupling that becomes increasingly relevant at low $n_\mathrm{2D}$.  The effect of this is to mix the different orbital symmetries.  While this is, in principle, straightforward to include in our calculations, it complicates their numerical solution greatly (Appendix~\ref{app:C}).  As we will show in the next section, however, the effects of the FE QCP do not depend greatly on details of the band structure, and we expect our main results to be robust.

\section{Results}
\label{sec:results}

\begin{figure}[tb]
  \includegraphics[width=\columnwidth]{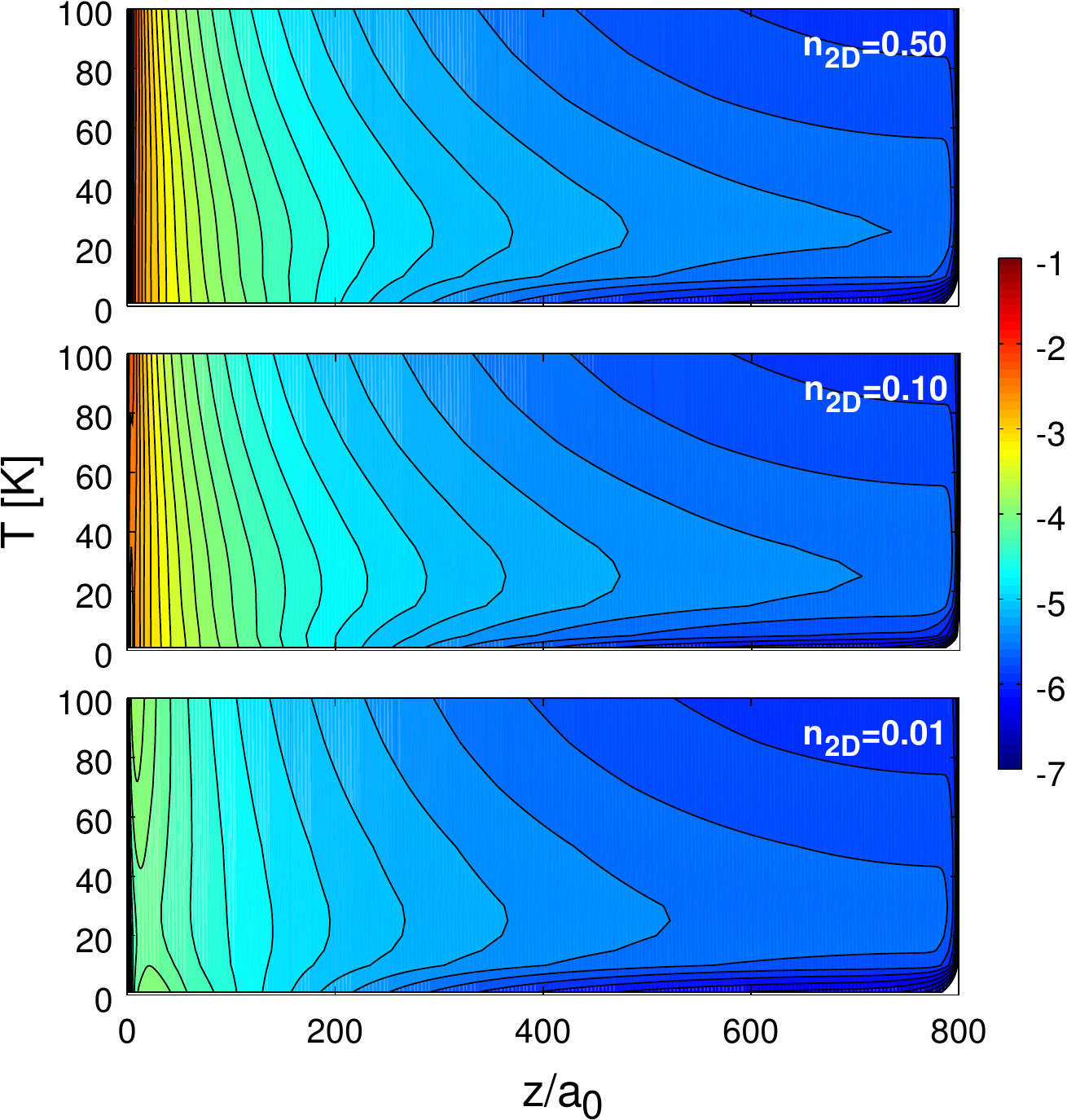}
\caption{Contour plot of $\log_{10} n(z,T) $ for (top) $n_\mathrm{2D} = 0.5$,
(middle) $n_\mathrm{2D} = 0.1$, and (bottom) $n_\mathrm{2D}=0.01$ electrons per 2D unit cell.  }
\label{fig:main_results}
\end{figure}

Figure~\ref{fig:main_results} shows the main numerical results for this work.  The calculated charge density $n(z)$ for STO slabs of thickness $L=800$ unit cells is plotted a range of 2D charge density values.  The largest electron density, $n_\mathrm{2D} = 0.5$ corresponds to the  half-electron per unit cell predicted by polar catastrophe models of interface doping.  In practice, values closer to $n_\mathrm{2D} = 0.1$ electron per unit cell are obtained, corresponding to the middle panel in Fig.~\ref{fig:main_results}.  The final set of results is for $n_\mathrm{2D} = 0.01$, which is close to where the metal-insulator transition is observed.  

At high and medium densities, $n(z)$ can approximately be decomposed into an interfacial component, localized to less than $\sim 30$ unit cells of the interface, and a long tail.   The interfacial component falls off with $z$ approximately as an exponential on a distance scale that is set by the shape of the confining electrostatic potential, while the tail falls off algebraically.\cite{Son:2009wb,Reich:2015ut,Gariglio:2015jx,Raslan:2016}  The total charge in the interface region is strongly doping-dependent, and at $n_\mathrm{2D} = 0.5$, more than 80\% of the charge lies within 10 unit cells of the interface, while that number falls to less than 15\% at $n_\mathrm{2D} = 0.01$.  This is evident in Fig.~\ref{fig:main_results}, where the intense interfacial peaks that are evident at $n_\mathrm{2D} = 0.5$ and 0.1 are gone at $n_\mathrm{2D} = 0.01$.  Conversely,  the tail component of the charge density varies only gradually with $n_\mathrm{2D}$ and has a universal shape.   A particularly striking feature of the tails is that $n(z)$ has a nonmonotonic temperature dependence that is peaked near $T=20$ K.  

The purpose of this work is to show that the universal structure of the tails is intimately tied to the quantum critical behaviour of the soft FE phonon mode.   
%In particular, we find that in the quantum critical regime, the long tails of the charge distribution satisfy
%\begin{equation}
%\frac{n(z,T)}{n_\mathrm{2D}} \sim \frac{\lambda(T)}{[z+\lambda(T)]^2},
%\end{equation}
%where $\lambda(T)$ is a temperature-dependent length scale that is proportional to $T \xi^2(T)$, where $\xi(T)$ is the FE correlation length.  The  non-monotonic $T$-dependence in Fig.~\ref{fig:main_results} tracks the temperature dependence of $T \xi^2(T)$.
To interpret our numerical results, we use a continuum single-band model that captures the essential elements of the more-involved numerical calculations.  With this approach, there are three constituent equations: the Schr\"odinger equation for the electronic wavefunctions;  Gauss' law,
\begin{equation}
\epsilon_\infty \frac{\partial E}{\partial z} = -en(z) - \frac{\partial P}{\partial z}, 
\label{eq:max1}
\end{equation}
for the electric field, subject to the boundary condition $E(0) = en_\mathrm{2D}/\epsilon_\infty a_0^2$ at the interface; and the constituent equation for the dielectric 
\begin{equation}
\left ( -\frac{d^2}{dz^2} + \xi^{-2} \right ) P(z) = \epsilon_\infty \xi_0^{-2} E(z),
\label{eq:Pz}
\end{equation}
which can be derived from the quantum Hamiltonian (\ref{eq:H}) within self-consistent phonon theory (Appendix~\ref{app:B}), and which is subject to the boundary conditions $P(0)=P(L) = 0$.  In Eq.~(\ref{eq:Pz}),  $\xi_0\sim 1$ \AA {} is a microscopic length scale and $\xi$ is the FE correlation length, which diverges at the QCP.  

Near the QCP, the correlation length  is a function of both temperature and polarization (Appendix~\ref{app:B}),
\begin{equation}
\xi^{-2}(T,P) = \xi^{-2}(0) +  {\cal A} T^{2\nu} + {\cal B} P^{\delta-1},
\label{eq:xim2}
\end{equation}
where $\nu$ and $\delta$ are critical exponents and ${\cal A}$ and ${\cal B}$ are constants.  Within self-consistent phonon theory, $\delta = 3$ and $2\nu = d/\bz-1$, where $d=3$ is the spatial dimension and $\bz=1$ is the dynamical critical exponent (Appendix~\ref{app:B}).  While the value  $\nu=1$ has been verified experimentally,\cite{Dec:2005cr,Rowley:2014bda} it has been reported\cite{Dec:1999wh,Prosandeev:1999tx} that $\delta = 2$.  This discrepency has, to our knowledge, not been explained.  The term $\xi^{-2}(0)$ in Eq.~(\ref{eq:xim2}) contains the effects of quantum fluctuations, and, as discussed above, can be made to vanish by oxygen isotope substitution. 

\begin{figure}
\includegraphics[width=\columnwidth]{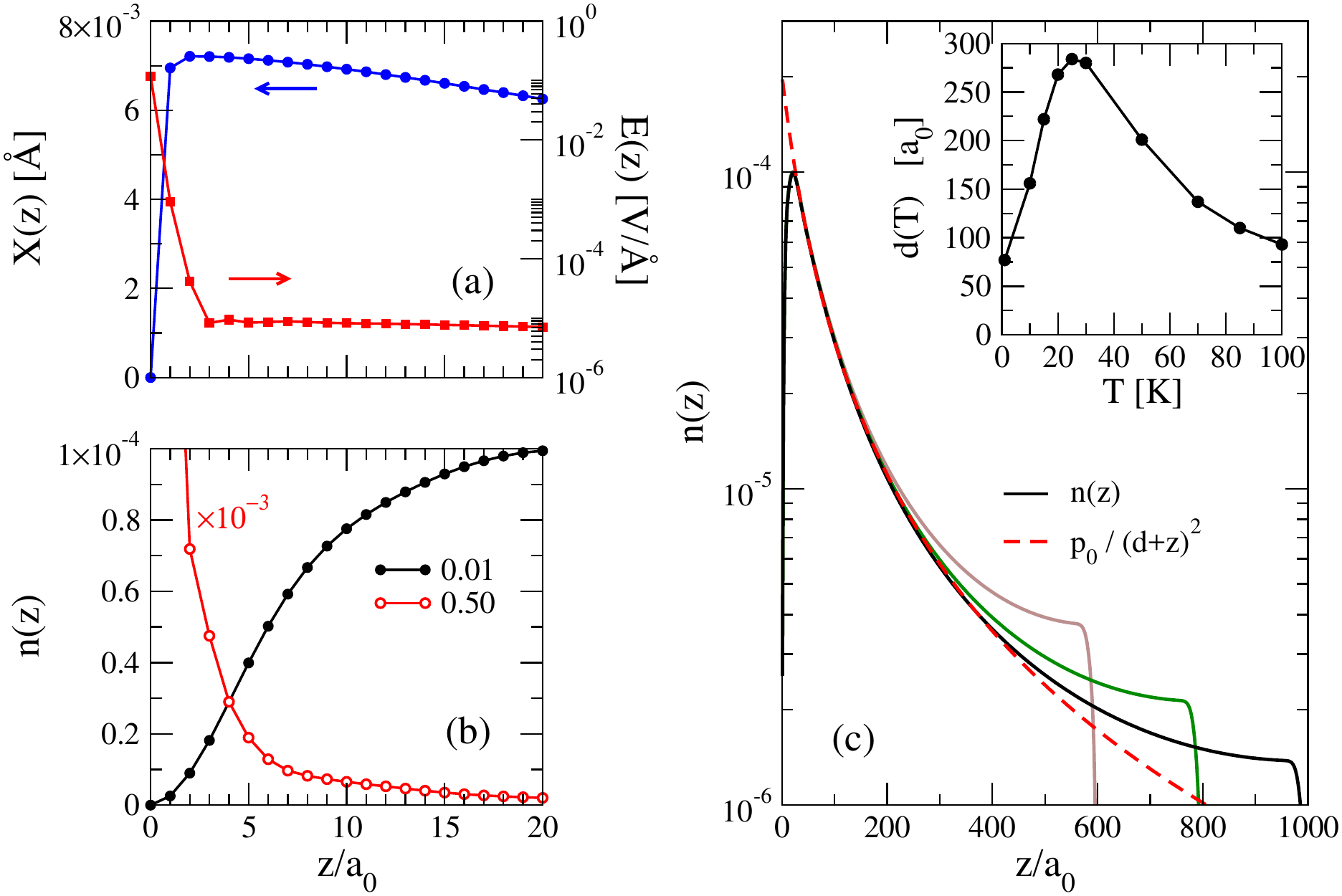}
\caption{Results of numerical calculations for $n_\mathrm{2D}=0.01$.  (a) Electric field $E(z)$ and normal coordinate $X(z)$ for the lattice polarization at $T=10$~K.  The polarization can be obtained from $P(z) = Q X(z)/a_0^3$, where $Q$ is an effective charge associated with the FE phonon mode. (b) Electron density near the interface, in units of electrons per unit cell. Results are shown for $n_\mathrm{2D} = 0.01$ and $n_\mathrm{2D} = 0.50$ electrons per 2D unit cell, at $T=10$~K.  Note that $n(z)$ is multiplied by $10^{-3}$ for $n_\mathrm{2D} = 0.50$.
(c) Electron density on a logarithmic plot emphasizing the long tails.  Results are shown for three different thicknesses of STO films,  $L = 600 a_0$, $800a_0$, and $1000 a_0$, illustrating finite size effects.  A fit of the $L=1000a_0$ data to $p_0/(\lambda +z)^2$  [following Eq.(\ref{eq:nzsol})] with $p_0 = 0.75a_0^2$ and $d = 61 a_0$ is shown.  At other temperatures, an additional exponential component $p_1\exp(-z/d_2)$ with $d_2 \sim 30a_0$ is required to obtain a good fit to the data.   {\it Inset:}  Temperature dependence of the length scale $\lambda(T)$.   }
\label{fig:orbits}
\end{figure}

Figure~\ref{fig:orbits} shows details of the numerical calculations for the low doping case, and we will use these as a guide for our analytic solution.  To begin, we focus on  the region near the interface.  At low doping, the confining electric field is weak and the electron density in the layers adjacent to the interface is small, as shown in Fig.~\ref{fig:orbits}(b).  On the length scale $z\sim \xi_0$, therefore, we can set $n(z)$ to zero in Eq.~(\ref{eq:max1}); making use of the boundary conditions at $z=0$, we then obtain $\epsilon_\infty E(z) = \epsilon_\infty E(0) - P(z)$.  Then,
Eq.~(\ref{eq:Pz}) becomes 
\begin{equation}
\left [-\partial_z^2 + \xi^{-2} + \xi_0^{-2} \right ]P(z) = \epsilon_\infty \xi_0^{-2} E(0).
%\left [-\partial_z^2 + \xi_L^{-2} \right ]P(z) = \epsilon_\infty \xi_0^{-2} E(0),
\label{eq:Pz2}
\end{equation}
Starting from the Lyddane-Sachs-Teller relationship between longitudinal and transverse phonon frequencies, we can identify  $\xi_L \equiv [\xi_0^{-2} + \xi^{-2}]^{-1/2}$  as the correlation length for the longitudinal polarization of the FE mode (Appendix~\ref{app:BB}).  The appearance of $\xi_L$ as the relevant length scale is expected because the field $E(z)$ is longitudinal. 
Because the length scale is $\xi_L$, and not $\xi$, the behavior in this region is noncritical.

 Solving Eqs.~(\ref{eq:Pz2}) and (\ref{eq:max1}), we then obtain
\begin{eqnarray}
P(z) &=& \epsilon_\infty E(0)\frac{\xi_L^2}{\xi_0^2} \left [1-e^{-z/\xi_L} \right ] \\ 
 E(z) &=& E(0) \xi_L^2 \left [ \xi^{-2} + \xi_0^{-2} e^{-z/\xi_L} \right ].
 \end{eqnarray} 
 This describes the relaxation of the dielectric, and the concommitant screening of the electric field, that occurs over the first few unit cells next to the interface, as shown in Fig.~\ref{fig:orbits}(a).  Because of this strong screening, 2D quantum well states are unable to form at the interface at low $n_\mathrm{2D}$.\cite{Raslan:2016}
 
 We emphasize that this situation is quite different from higher dopings, where the nonlinear term ${\cal B} P^{\delta-1}$ in Eq.~(\ref{eq:xim2}) limits the dielectric screening of the external field;  in this case, the external field creates a deep quantum well that confines the majority of the charge to within $\sim 10$ unit cells of the interface [Figs.~\ref{fig:main_results} and \ref{fig:orbits}(b)].  This 2DEG comprises a number of quantum well states with strongly 2D character.  The interface region is still noncritical, however,  because of the strong interfacial electric field.

Next, we focus on the tail region of the STO slab. Because of the  screening of the electric field, either by the dielectric at low $n_\mathrm{2D}$ or by the electron gas at high $n_\mathrm{2D}$, the electrostatic potential is slowly varying in the tails, and we can make a local (Thomas-Fermi) approximation for the charge density.  Numerically, we find that the electrochemical potential $\mu$ lies below the electrostatic potential $-eV(z)$ in the tails at all temperatures, so that even at the lowest temperature  we have studied ($T=1$K), the Fermi-Dirac distribution can be treated in a high temperature approximation.  Then, the charge density is 
\begin{eqnarray}
n(z) &=& \frac{2}{4\pi^2} \left( \frac{2m^\ast}{\hbar^2} \right )^{3/2} \int_{0}^\infty \sqrt{\epsilon} f[\epsilon-eV(z)] d\epsilon \nonumber \\
&\approx& 2 \left( \frac{m^\ast}{2\pi \hbar^2 \beta} \right )^{3/2}
e^{\beta[\mu+eV(z)]},
\label{eq:nz}
\end{eqnarray}
where $f(\epsilon)$ is the Fermi function, and $\beta = 1/k_BT$.  Equation~(\ref{eq:nz}) can equivalently be expressed as a differential equation
\begin{equation}
\frac{\partial n}{\partial z} = -\beta e n(z) E(z).
\label{eq:nz_diff}
\end{equation} 

In the tails, we find that $\partial^2P/\partial z^2 \ll P/\xi^2$, so that Eq.~(\ref{eq:Pz}) simplifies to $\epsilon_\infty E(z) = (\xi_0^2/\xi^2) P(z)$.  Because $\xi_0/\xi \ll 1$ near the QCP, it follows that the electric field inside the STO is extremely weak, and to a good approximation Eq.~(\ref{eq:max1}) reduces to $-en(z) = \partial_z P(z)$.   This point is key, as it is a statement that  electric fields associated with the longitudinal component of the polarization are screened by the conduction electrons.  These electric fields are responsible for the hardening of the longitudinal phonon mode relative to the transverse mode, as expressed by the Lyddane-Sachs-Teller relationship; their absence implies that the longitudinal dielectric response, and therefore the charge density in the tails, exhibits quantum criticality.  

This is seen directly by solving Eqs.~(\ref{eq:max1}), (\ref{eq:Pz}) and (\ref{eq:nz_diff}) using the two simplifications given above:\cite{GouyChapman}
\begin{eqnarray}
P(z) &=& en_\mathrm{2D}\frac{\lambda(T)}{\lambda(T) + z}, \label{eq:Pzsol} \\
E(z) &=& \frac{en_\mathrm{2D}}{\epsilon_\infty}\frac{\xi_0^2}{\xi^2} \frac{\lambda(T)}{\lambda(T)+z}, \label{eq:Ezsol} \\
n(z) &\approx& -\frac{1}{e}\partial_zP(z) = n_\mathrm{2D} \frac{\lambda(T)}{[\lambda(T) + z]^2}, \label{eq:nzsol}
\end{eqnarray}
where $\lambda(T)$ is determined by the normalization condition $\int_0^\infty n(z) dz = n_\mathrm{2D}$, from which 
\begin{equation}
\lambda(T) = \frac{\epsilon_\infty 2k_BT \xi^2(T) }{n_\mathrm{2D} e^2 \xi_0^2}.
\label{eq:lambda}
\end{equation}  
This expression holds at low $n_\mathrm{2D}$, where the exponentially confined component of the 2DEG can be ignored;  when the interfacial component of the electron density is significant, $n_\mathrm{2D}$ in this expression must be replaced by the total charge density in the tail.

Figure~\ref{fig:orbits} shows fits of Eq.~(\ref{eq:nzsol}) to numerical results for the lattice model.  In general, we find that the decaying part of the charge density can be fitted to the sum of an inverse quadratic term, as in Eq.~(\ref{eq:nzsol}), and an exponential term $p_1 \exp(-z/d_2)$, where $d_2 \sim 30 a_0$, representing the 2D interfacial component.   The model parameters $n_\mathrm{2D} = 0.01$ and $T=10$ K in Fig.~\ref{fig:orbits} are chosen such that the exponential component has nearly vanishing weight.  Indeed, in Fig.~\ref{fig:main_results} this corresponds to the temperature and doping where the charge density is most spread out.

The length scale $\lambda(T)$ sets the range over which the charge density falls off.  
The temperature-dependence of $\lambda(T)$ is given by the product $T \xi^2(T)$ in Eq.~(\ref{eq:nzsol}).  In the tail region, the polarization is weak and the nonlinear  term $P^{\delta-1}$ can be set to zero in Eq.~(\ref{eq:xim2}). We then arrive at the form
\begin{equation}
\lambda(T) \sim \frac{T}{ \xi^{-2}(0) + {\cal A} T^{2\nu}},
\label{eq:dT}
\end{equation}
which is a maximum at  $T^{2\nu} =  \xi^{-2}(0)/(2\nu-1){\cal A}$.     This equation is  consistent with the nonmonotonic charge distribution shown in Fig.~\ref{fig:main_results}:  for $T \lesssim 25$~K, the range of the tail grows linearly with $T$, while it falls as $1/T$ for $T \gtrsim 25$~K.  This is made explicit in Figure~\ref{fig:orbits}(c), which shows  $\lambda(T)$ obtained by fitting to the results shown in Fig.~\ref{fig:main_results}(c).

\section{Discussion}
Equations~(\ref{eq:Pzsol})-(\ref{eq:nzsol}), along with Eq.~(\ref{eq:dT}), constitute the main results of this work.  They exhibit typical quantum critical behavior, namely that $E(z)$, $P(z)$ and $n(z)$ fall off algebraically with distance.  Equation~(\ref{eq:dT}) in particular shows that the length scale $\lambda(T)$ is connected to the quantum critical properties of STO.  Quantum critical behavior dominates when  ${\cal A} T^{2\nu} > \xi^{-2}(0)$, and in this region
\begin{equation}
\lambda(T) \sim T^{1-2\nu} = T^{2-d/\bz}.
\end{equation}
Because $\xi^{-2}(0)$ depends  on the atomic masses, the crossover temperature can be tuned downwards by isotope substitution, and will vanish at the QCP.   At the QCP, $\lambda(T=0)$ diverges and the charge spreads uniformly into the STO substrate.

At higher temperatures, there is a second crossover to noncritical Curie-like behavior, namely $\xi^{-2} \sim T$.    Within self-consistent phonon theory, the fluctuations cross over to the classical limit when $T \approx \hbar\tilde \omega/2\pi k_B$, where $\tilde \omega = v_\mathrm{ph}/a_0$ is a typical phonon frequency.\cite{Palova:2009js} [This result can be obtained by analyzing the temperature-dependence of Eq.~(\ref{eq:dX2B}).]  In this regime, $n(z)$ is still given by Eq.~(\ref{eq:nzsol}), but with a temperature-independent $\lambda$.  Experimentally, deviations from quantum critical scaling appear at $T \gtrsim 35$~K and Curie-like behavior is found above $\sim 100$~K.\cite{Dec:1999wh}  Our calculations, which are based on fits to the low-temperature susceptibility, overestimate the crossover temperature somewhat. 

 The progressive crossover between 2D and 3D that occurs as $n_\mathrm{2D}$ is lowered has implications for the metal-insulator transition that has been observed  at $n_\mathrm{2D} \sim 0.01 a_0^{-2}$ ($\sim 10^{13}$ cm$^{-2}$).\cite{Thiel:2006eo,Liao:2011bk,Pallecchi:2015et}  It is commonly assumed that this transition corresponds to a localization of electrons belonging to a single 2D interface state, and indeed there is some experimental evidence suggesting that a single band is occupied at low densities.\cite{Joshua:2012bl} Alternatively, our calculations suggest that at low $n_{2D}$ a dilute electron gas spreads away from the interface;  at low temperatures, the electron gas may be trapped by crystal defects, leading to an insulating state.  For $n_\mathrm{2D}=0.01$, the maximum charge density at 10~K   is $n_\mathrm{max} \sim 10^{-4}$ per unit cell, corresponding to $\sim 10^{18}$ cm$^{-3}$ [Fig.~\ref{fig:main_results}(c)].  For comparison, high-quality single crystals of STO become insulating below electron densities of $n_c \sim 10^{16}$--$10^{17}$ cm$^{-3}$,\cite{Spinelli:2010dm} which is an order of magnitude lower than $n_\mathrm{max}$.  It is possible, however, that $n_c$ is higher near interfaces than in bulk crystals because of defects introduced during interface growth.

Finally, we discuss recent analytic calculations of the electron density performed by Reich {\em et al,}\cite{Reich:2015ut}  who also predicted an algebraic decay of the electron density away from STO interfaces.  They found that $n(z) \sim (\lambda+z)^{-6}$ in the linear regime (weak electric fields) and $n(z) \sim (\lambda+z)^{-12/7}$ when the dielectric response is nonlinear (strong electric fields), and argue that the latter case is consistent with experiments.  Several calculations have shown that nonlinear effects are important at large doping, but that the response is linear below a crossover doping $n_\mathrm{2D} \sim 10^{14}$cm$^{-2}$ ($0.16$ electrons per 2D unit cell).\cite{Copie:2009ev,Khalsa:2012fu,Gariglio:2015jx}  Based on this, one expects the novel $12/7$ power law to apply at high densities and that at low densities, the density should fall off extremely rapidly as $(\lambda+z)^{-6}$; in contrast, we find a universal quadratic tail at all dopings.

There are two main reasons that our calculations differ from Ref.~\onlinecite{Reich:2015ut}.  First, electric fields in our calculations are strongly screened by the interfacial component of the 2DEG so that the tails are always in the linear regime even if the interface region is not.\cite{Raslan:2016}  Second, Ref.~\onlinecite{Reich:2015ut} makes the standard assumption that at sufficiently low temperatures the electron density can be obtained from the zero-temperature limit.  Conversely, we find that the electrochemical potential always lies slightly below the bottom of the conduction band in the tail region, such that the electron distribution must always be treated in the high-temperature limit, even at $T=1$K.  This leads directly to Eq.~(\ref{eq:nz_diff}) for the electron density, rather than the more usual zero-temperature result, $n(z) \propto [\epsilon_F + eV(z)]^{3/2}$.

\section{Conclusions}
We have studied a model SrTiO$_3$ interface in which both the electron and phonon degrees of freedom are treated quantum mechanically.  We calculated the charge distribution near the interface numerically, and showed analytically that the profile of the charge distribution is shaped by the quantum critical behavior of the soft ferroelectric phonon mode.   At all electron densities $n_\mathrm{2D}$, we find universal tails that extend far into the SrTiO$_3$ substrate. At high $n_\mathrm{2D}$ these tails are masked by prominent interface states that contain the majority of the conduction electrons; at low $n_\mathrm{2D}$, however, the majority of the charge lies in the tails.  We speculate that the metal-insulator transition that is observed at low $n_\mathrm{2D}$ corresponds to charge trapping of the dilute electron gas in the tails.

\section*{Acknowledgments}
We thank B.\ I.\ Shklovskii for pointing out the connection between our work and the Gouy-Chapman theory of screening in ionic liquids.
This work has been supported by the Natural Sciences and Engineering Research 
Council (NSERC) of Canada.

\appendix
\section{Self-Consistent Phonon Theory}
\label{app:A}

Self-consistent phonon theory is based on the mean-field decomposition
\begin{equation}
\left ( \sum_{\alpha} \hat X_{i\alpha}^2 \right )^2 \approx 2 \sum_{\alpha\beta} \hat X_{i\alpha}^2
 \langle \hat X_{i\beta}^2\rangle - \left ( \sum_{\alpha} \langle \hat X_{i\alpha}^2 \rangle \right )^2.
\end{equation}
The mean-field Hamiltonian is then
\begin{eqnarray}
\hat H_\mathrm{scp} &=& \sum_{i\alpha} \frac{\hat \Pi_{i\alpha}^2}{2M} 
+ \frac 12 \sum_{ij\alpha} \hat X_{i\alpha} D_{ij}^{\alpha} \hat X_{j\alpha}  \nonumber \\
&& - Q\sum_{i\alpha} E_{i\alpha} \hat X_{i\alpha} 
-\frac{B}{4\eta} \sum_i \left ( \sum_{\alpha} \langle \hat X_{i\alpha}^2 \rangle \right )^2 
\label{eq:H1}
\end{eqnarray}
where
\begin{eqnarray}
D_{ij}^\alpha &=&  D_{ij}^{0\alpha} + \delta_{i,j} \frac{B}{\eta}  \sum_{\beta} \langle \hat X_{i\beta}^2 \rangle, \nonumber \\
  &=& \tilde D_{ij}^{0\alpha} + \delta_{i,j} \frac{B}{\eta}  \sum_{\beta} \left ( \langle \hat X_{i\beta}^2 \rangle
  -  \langle \hat X_{i\beta}^2 \rangle_0 \right )
\label{eq:D}
\end{eqnarray}
are the renormalized elastic constants, and $\delta_{i,j}$ is the Kronecker delta function. We have defined 
\begin{equation}
\tilde D_{ij}^{0\alpha} = D_{ij}^{0\alpha} + \delta_{i,j}\frac{B}{\eta}\sum_\beta \langle \hat X_{i\beta}^2 \rangle_0,
\label{eq:Dtilde}
\end{equation} 
where $\langle \hat X_{i\beta}^2 \rangle_0$ describes quantum fluctuations at zero temperature and electric field.  We take a nearest-neighbor model for $\tilde D^{0\alpha}_{ij}$ with
\begin{equation}
\tilde D^{0\alpha}_{ij} = (A+12C) \delta_{i,j} -  2C \delta_{\langle i,j\rangle}, 
\label{eq:D0}
\end{equation}
where $\delta_{\langle i,j\rangle}$ is 1 if $i$ and $j$ are nearest-neighbors and 0 otherwise.  The parameter $A$ contains both the classical deformation potential $D_{ii}^{0\alpha}$ and the quantum flucuation term in Eq.~(\ref{eq:Dtilde}).

The 3D Fourier transform of Eq~(\ref{eq:D0}) gives the FE phonon dispersion for bulk STO zero temperature and electric field:
\begin{equation}
M\omega_\bq^2 = A - 4C \left [\cos (q_xa_0) + \cos (q_ya_0) +\cos (q_za_0) - 3 \right].
\end{equation}
The parameter $A$ is negative in the absence of quantum fluctuations, such that $\omega_\bq$ is imaginary at the Brillouin zone centre.  This signals a FE instability; however, the quantum fluctuation term in Eq.~(\ref{eq:Dtilde}) is positive and sufficiently large such that $A$ is positive for ST$^{16}$O.  This is the origin of quantum paraelectricity.

The fluctuation term $\langle \hat X_{i\beta}^2\rangle$ in Eq.~(\ref{eq:D}) must be
found self-consistently.  For this purpose, it is convenient to re-cast $\hat H_\mathrm{scp}$ as
\begin{equation}
\hat H_\mathrm{scp} = \sum_{i\alpha} \frac{\hat \Pi_{i\alpha}^2}{2M} 
+ \frac 12 \sum_{ij\alpha} (\hat X_{i\alpha} - X_{i\alpha} ) D_{ij}^{\alpha} (\hat X_{j\alpha}
-X_{j\alpha})
% \nonumber \\
%&&+ \frac 12 \sum_{ij\alpha}  X_{i\alpha} D_{ij}^{\alpha} X_{j\alpha}
\label{eq:H2}
\end{equation}
where the constant terms, which do not influence lattice dynamics, have been dropped.  $X_{i\alpha}$ is the shift in the equilibrium normal coordinate due to the electric field ${\bf E}$:
\begin{equation}
X_{i\alpha} = Q \sum_j [{\bf D}^\alpha]^{-1}_{ij} E_{j\alpha},
\label{eq:X}
\end{equation}
where $[\ldots]^{-1}$ is a matrix inverse. It follows from the symmetry of Eq.~(\ref{eq:H2}) that $\langle \hat X_{i\alpha} \rangle = X_{i\alpha}$. 

$\hat H_\mathrm{scp}$ in Eq.~(\ref{eq:H2}) has a quadratic form and can by solved by numerically diagonalizing ${\bf D}^\alpha$.
 It is then straightforward to show that
\begin{equation}
\langle \hat X_{i\alpha}^2 \rangle = \langle ( \hat X_{i\alpha} - X_{i\alpha})^2\rangle
+ X_{i\alpha}^2,
\label{eq:X2}
\end{equation}
with 
\begin{equation}
\langle ( \hat X_{i\alpha} - X_{i\alpha})^2\rangle = \frac{\hbar}{2M} \sum_n  \frac{|S_{i,n}^\alpha|^2}{\omega_{n\alpha}} \coth\left( \frac{\hbar \omega_{n\alpha}}{2k_BT} \right ).
\label{eq:dX2}
\end{equation}
In this expression, 
$M \omega_{n\alpha}^{2}$ is the $n$th eigenvalue of ${\bf D}^\alpha$, and $S_{:,n}^\alpha$ is the corresponding eigenvector (the subscript ``:'' stands for an entire row or column of a matrix).

In summary, the full solution of the phonon spectrum requires self-consistently determining the mean normal mode displacements $X_{i\alpha}$ [Eq.~(\ref{eq:X})] and fluctuations $\langle ( \hat X_{i\alpha} - X_{i\alpha})^2\rangle$ [Eq.~(\ref{eq:X2})].   For the slab geometry, these equations may be simplified by making use of the translational invariance in the $x$-$y$ directions.  Then, Eq.~(\ref{eq:X}) becomes
\begin{equation}
X_{i_z z} =  Q\sum_{j_z} [{{\bf D}^{z}_{\bq=0}}^{-1}]_{i_zj_z} E_{j_z z}.
\label{eq:X2D}
\end{equation}
Here, ${\bf D}^\alpha_\bq$ has been Fourier transformed along the $x$ and $y$ directions, so
the unit cell coordinates $(i_x,i_y,i_z)$ transform to $(\bq,i_z)$ with $\bq = (q_x,q_y)$;
${\bf D}^\alpha_\bq$ is thus an $L\times L$ matrix  in terms of layer index:
\begin{equation}
{\bf D}^\alpha_\bq = {\bf D}^0 + (A  -4C[F_2(\bq)-1]) {\bf 1},
\end{equation}
 where 
$F_2(\bq) = \cos q_x + \cos q_y - 2$,
${\bf 1}$ is the $L\times L$ identity matrix, and 
\begin{equation}
{\bf D}^0 = \left [ \begin{array}{ccccc}
B\delta X^2_1 & -2C & 0 & \ldots & \\
-2C & B\delta X^2_2 & -2C & \\
&&\ddots \\
&&&B\delta X^2_{L-1} & -2C \\
&&&-2C &B\delta X^2_L
\end{array}
\right ],
\end{equation}
where 
\begin{equation}
\delta X^2_{j_z} = \frac{B}{\eta}\sum_{\alpha=1}^\eta \langle (\hat X_{i_z\alpha} - X_{i_z\alpha})^2 \rangle + \frac {B}{\eta} X_{i_z z}^2.
\end{equation}
%\begin{eqnarray}
%[{\bf D}^\alpha_{\bq}]_{i_zj_z} &=& \delta_{i_z,j_z} \Big \{ A  -4C[F_2(\bq)-1]  \nonumber \\
%&& +  \frac{B}{d}\sum_{\beta} \left ( \langle \hat X^2_{i_z \beta}\rangle - \langle \hat X^2_{i_z   \beta}\rangle_0 \right ) 
% \Big \} \nonumber \\
% &&-\delta_{\langle i_z,j_z \rangle} 2C,
 %\label{eq:D2}
%\end{eqnarray}

In the slab geometry, the fluctuation equation (\ref{eq:X2})  becomes
\begin{equation}
\langle \hat X_{i_z\beta}^2\rangle 
=\frac{\hbar}{2M N_q} \sum_{\bq n} [{\bf v}_{\bq\beta}^n]_{i_z}^2 \frac{ \coth\left( \frac{\hbar \omega_{\bq n\beta}}{2k_BT} \right )
}{\omega_{\bq n\beta}} + \delta_{\beta,z} X_{i_z z}^2,
\label{eq:X2qxy}
\end{equation}
where $N_q$ is the number of $\bq$-points in the sum and
where ${\bf v}_{\bq\beta}^n$ is the eigenvector of ${\bf D}^\alpha_\bq$,
\begin{equation}
{\bf D}^\alpha_\bq {\bf v}^{n}_{\bq \alpha}  = M\omega_{n\bq\alpha}^2 {\bf v}_{\bq\alpha}^n,
\end{equation}
describing the $n$th phonon eigenmode with eigenfrequency $\omega_{n\bq\alpha}$ of the layered system.  In this expression, $n \in [1,L]$ refers to the phonon band and $\bq$ the 2D phonon wavevector.    

Because of the simple structure of ${\bf D}_\bq^\alpha$, ${\bf v}^n_{\bq \alpha}$ are also eigenvectors of ${\bf D}^0$, and are therefore independent of $\bq$.  The phonon eigenfrequencies then satisfy
\begin{equation}
M\omega_{n\bq\alpha}^2 = M\omega_{n\bq=0\,\alpha}^2 - 4CF_2(\bq).
\end{equation}
It is thus only necessary to perform a single numerical diagonalization, rather than for each value of $\bq$.

\section{Relation of the Quantum Phonon Model to the Continuum Model}
\label{app:B}
There are several ways to obtain the continuum model for the polarization, Eq.~(\ref{eq:Pz}).  The most direct starts with Eq.~(\ref{eq:D0}).  For the slab geometry, in which both the polarization and its gradient are along the $z$ axis, we recognize that 
\begin{equation}
12C\delta_{i,j} - 2C\delta_{\langle i,j \rangle} \approx - 2Ca_0^2 \nabla^2 \rightarrow -2Ca_0^2 \frac{\partial^2}{\partial z^2}
\end{equation}
Then Eq.~(\ref{eq:X}) can be recast as $QE_{i_z} = \sum_{j_z} D^z_{i_zj_z}  X_{j_z z}$, or
\begin{equation}
QE(z) = \left [- 2Ca_0^2 \frac{\partial^2 }{\partial z^2} + \left (A +
B\delta X^2(z) \right ) \right ] X(z),
\end{equation}
where we have taken $X_{j_z z} \rightarrow X(z)$.
Dividing through by $2Ca_0^2$ and multiplying by $Q/a_0^3$ gives Eq.~(\ref{eq:Pz}) with
\begin{eqnarray}
\xi_0^{-2} &=& \frac{Q^2}{\epsilon_\infty Mv_\mathrm{ph}^2 a_0^3}, \label{eq:x0m2} \\
\xi^{-2} &=&  \frac{1}{M v_\mathrm{ph}^2} \left (A + B\delta X^2 + \frac {B}{\eta} X^2 \right ),
\end{eqnarray}
where $v_\mathrm{ph} = \sqrt{2C/M}a_0$ is the phonon velocity at the QCP.  Away from the interface, $\delta X^2$ is nearly independent of $z$, so that $\xi^{-2}$ can be treated as a constant.  Using the parameters from Table~\ref{table:1}, we estimate $\xi_0 \sim 1$ \AA{} and  $\xi(T=0,E=0) \sim 54$ \AA.  

Finally, we can find the leading-order temperature- and electric field-dependence, which is relevant to the tail region of the charge density.  In a $d$-dimensional crystal, Eq.~(\ref{eq:dX2}) can be Fourier transformed to obtain
\begin{eqnarray}
\delta X^2 &=& \frac{\hbar}{2M N_q} \sum_{\bq} \frac{1}{\omega_{\bq}} \left[ \coth\left( \frac{\hbar \omega_{\bq}}{2k_BT} \right ) -1 \right]
\label{eq:dX2B}
\end{eqnarray}
where $\bq$ is a $d$-dimensional wavevector.  At the QCP, the soft phonon mode is gapless.  For the general case, we can let $\omega_\bq = \overline \omega (a_0 q)^\bz$, where $\overline \omega$ sets the width of the dispersion and $\bz$ is the dynamical critical exponent.  We then obtain $\xi^{-2} = \xi^{-2}(0) + {\cal A} T^{d/\bz-1} + {\cal B} P^2$.  For the specific case of $d=3$ and $\bz=1$, as appropriate for bulk STO, we obtain
\begin{eqnarray}
\xi^{-2}(0) &=& \frac{A}{Mv_\mathrm{ph}^2}, \\
%{\cal A} &=& \frac{k_B^2 B}{M \pi^2 v_\mathrm{ph}^2} \int_0^\infty y \left[ \coth(y)-1\right] dy. \\
{\cal A} &=& \frac{B k_B^2 a_0^3}{\hbar \pi^2 M^2 v_\mathrm{ph}^5}  \int_0^\infty y \left[ \coth(y)-1\right] dy, \\
{\cal B} &=& \frac{B a_0^6}{\eta M v_\mathrm{ph}^2  Q^2}.
\label{eq:A}
\end{eqnarray}

\section{Relationship between longitudinal and transverse phonon correlation lengths}
\label{app:BB}
The Lyddane-Sachs-Teller equation relates the transverse and longitudinal frequencies of a particular phonon mode at $\bq=0$ via $\omega_L^2 = \frac{\epsilon}{\epsilon_\infty} \omega_{\bq=0}^2$, where $\omega_L$ is the longitudinal phonon frequency.  The total dielectric permittivity can be written as the sum of contributions from the atomic polarizability ($\epsilon_\infty$) and the lattice response,
\begin{equation}
\epsilon = \epsilon_\infty + \epsilon_0 \chi^\mathrm{lat.}_{\bq=0}
\label{eq:lst}
\end{equation}
where $\chi_{\bq=0}^\mathrm{lat.}$ is the dielectric susceptibility of the lattice.

In a bulk three-dimensional crystal, Eq.~(\ref{eq:X}) simplifies to $X_{\bq \alpha} = Q E_{\bq\alpha}/M\omega_{\bq\alpha}^2$, where $\bq$ is a 3D wavevector.  Taking the polarization to be $P_{\bq\alpha} = Q  X_{\bq \alpha}/a_0^3$, we obtain the lattice susceptibility
\begin{equation}
\chi_{\bq,\alpha\beta}^\mathrm{lat.} = \frac{1}{\epsilon_0} \frac{\partial P_{\bq\alpha}}{\partial E_{\bq\beta} }
= \frac{Q^2}{\epsilon_0 a_0^3 M \omega_{\bq\alpha}^2} \delta_{\alpha,\beta}.
\label{eq:chiq}
\end{equation}
In an isotropic material, $\chi_{\bq,\alpha\alpha}^\mathrm{lat.}$ is independent of the polarization direction $\alpha$.  

Substituting Eqs.~(\ref{eq:lst}) and (\ref{eq:chiq}) into the Lyddane-Sachs-Teller relation, we obtain
\begin{equation}
\omega_L^2 = \omega_{\bq=0}^2 + \frac{Q^2}{\epsilon_0 a_0^3 M}.
\end{equation}
Then, from Eq.~(\ref{eq:x0m2}), 
\begin{eqnarray}
\xi_L^{-2} \equiv \frac{\omega_L^2}{v_\mathrm{ph}^2} &=& \frac{\omega_{\bq=0}^2}{v_\mathrm{ph}^2}
+ \xi_0^{-2} \nonumber \\
&=& \xi^{-2} + \xi^{-2}_0.
\end{eqnarray}
This shows that the length scale that appears in Eq.~(\ref{eq:Pz2}) is associated with the longitudinally polarized phonon mode.

\section{Electronic Band Model}
\label{app:C}
%We consider the formation of a low density electronic gas at a SrTiO$_3$ surface.  We use a simplified discretization of both the charge and the polarization, as illustrated in Fig.~\ref{fig:structure}.  We imagine that the free electrons are confined to 2D TiO$_2$ sheets, labelled by indices $i_z = 1,\ldots,L$, separated by layers of dielectric with polarizations $P_{i_z}$ along the $z$ axis.  A cap layer at $i_z=0$ with a fixed 2D charge density $en^\mathrm{ext}/a^2$ generates a potential well that confines the charge to the interface.  We assume translational invariance in the $x$-$y$ plane, so that the total (screened) potential energy depends only on the layer index.  Assuming the charge to be homogeneously distributed within the TiO$_2$ layers, we obtain the potential energy
%\begin{equation}
%\phi_{j_z} = \frac{e^2}{2\epsilon_0 \epsilon_\infty a} \sum_{i_z} (|i_z-j_z|-i_z) (n^b_{i_z} - n^f_{i_z} + n^\mathrm{ext}\delta_{i_z,0}).
%\label{eq:phi}
%\end{equation}
%where $n^f_{i_z}$, $n^b_{i_z}$, and $n^\mathrm{ext}$ are, respectively, the number densities (per unit cell) of  the free electrons, the polarization charge, and the external charge; $\epsilon_\infty$ is the high frequency contribution to the dielectric constant, and $a$ is the lattice constant.  The form of the potential is chosen such that $\phi_{j_z}$ vanishes at $j_z = 0$.

  Because we neglect spin-orbit coupling, our Hamiltonian is block diagonal in the orbital type $\alpha$, so that electrons have pure $d_{xy}$, $d_{yz}$, or $d_{zx}$ character.  For the slab geometry, the Hamiltonian\cite{Raslan:2016} for orbital type $\alpha$ can be written as 
\begin{equation}
{\bf H}_\alpha (\bk) = {\bf H}_{0\alpha} + \epsilon_{\alpha\bk} {\bf 1}
\end{equation}
where $\bk = (k_x,k_y)$ are 2D wavevectors and ${\bf H}_\alpha(\bk)$ is an $L\times L$ matrix with rows and columns corresponding to the different layers in the STO slab, and 
\begin{equation}
\epsilon_{\alpha\bk} = -2t_{\alpha x} \cos(k_x a_0) - 2_{\alpha y}\cos(k_y a_0).
\end{equation}
Here, $t_{\alpha x}$ is either $t_\|$ or $t_\perp$, as appropriate for hopping in the $x$ direction for orbital type $\alpha$.  At low fillings, we write
\begin{equation}
\epsilon_{\alpha\bk} \approx -2(t_{\alpha x} + t_{\alpha y})+ \frac{\hbar^2}{2} \left( \frac{k_x^2}{m_{\alpha x}}+\frac{k_y^2}{m_{\alpha y}} \right ),
\end{equation}
where $m_{\alpha (x,y)} = [2 t_{\alpha (x,y)} a_0^2/\hbar^2]^{-1}$ 
are the effective masses along $x$ and $y$ directions.  These 2D dispersions
are coupled by interlayer hopping matrix elements $t_{\alpha z}$.  
%  For a given $\alpha$, the total Hamiltonian can be written as an $L\times L$ matrix in the layer index, of the form ${\bf H}_\alpha (\bk) = {\bf H}_{0\alpha} + \epsilon_{\alpha\bk} {\bf 1}$, where ${\bf 1}$ is the rank-$L$ identity matrix, and
\begin{equation}
{\bf H}_{0\alpha} = 
\left [ \begin{array}{ccccc}
-eV_1 & t_{\alpha z} & 0 & \ldots \\
t_{\alpha z} & -eV_2 & t_{\alpha z} & \\
&&\ddots \\
&&&-eV_{L-1}&t_{\alpha z} \\
&&&t_{\alpha z} &-eV_L
\end{array}\right ].
\label{eq:H0}
\end{equation}
The diagonal elements of ${\bf H}_{0\alpha}$ are obtained from Eq.~(\ref{eq:V}).

Because of its simple structure, ${\bf H}_\alpha(\bk)$ has common eigenvectors $\Psi^\alpha_{j_z n}$ with
${\bf H}_{0\alpha}$.  Furthermore, given eigenvalues $E_{\alpha n}$ of ${\bf H}_{0\alpha}$, the eigenvalues of ${\bf H}_\alpha(\bk)$ are trivially $E_{\alpha n\bk} = E_{n}^\alpha + \epsilon_{\alpha \bk}$. It then follows that the electron density (per 2D unit cell) in layer $j_z$ is  
\begin{eqnarray}
n_{j_z}^f &=& \frac{1}{N_2} \sum_{\bk} \sum_{\alpha n} |\Psi^\alpha_{j_z n}|^2 f(E_{\alpha n} + \epsilon_{\alpha \bk}) \nonumber \\
&=&\frac{\sqrt{m_{\alpha x} m_{\alpha y}} }{\pi \hbar^2}\sum_{\alpha n} |\Psi^\alpha_{jn}|^2  \int_0^\infty d\epsilon f(E_{\alpha n} + \epsilon)
\nonumber \\
&=&\frac{\sqrt{m_{\alpha x} m_{\alpha y}} }{\beta \pi \hbar^2}\sum_{n=1}^L \sum_{\substack{\alpha=xy,\\xz,yz}} |\Psi^\alpha_{jn}|^2 \ln\left ( 1+e^{\beta(\mu-E_{\alpha n})}
\right ) \nonumber \\
\label{eq:nf}
\end{eqnarray}
where $f(x)$ is the Fermi-Dirac distribution at $\beta = 1/k_BT$.

%\bibliography{SrTiO3,LaTiO3,magnetism,superconductivity,transport,bandstructure,other}
\bibliography{interfaces}

\begin{thebibliography}{41}
\bibitem [{Gou()}]{GouyChapman}%
  \BibitemOpen
  \href@noop {} {}\bibinfo {note} {This solution is essentially the same as
  that of Gouy and Chapman for the screening of a planar charge by a classical
  ionic liquid. See Ref.~\onlinecite{GrosbergRMP:2002} for a
  discussion}\BibitemShut {NoStop}%
\end{thebibliography}

\end{document}